\newcommand{\lb}{${\cal L}_{B}$}

\newcommand{\ie}{{\it i.e.}}
\newcommand{\eg}{{\it e.g.}}
\newcommand{\GRS}{{GRS 1915+105}}
\newcommand{\alf}{Alfv\'en }
\documentclass{PoS}

\title{Global MHD instabilities: from Low Frequency to High Frequency QPOs, and to Sgr A*}

\ShortTitle{LFQPO, HFQPO and Sgr A*}

\author{\speaker{Michel Tagger}\\
        Service d'Astrophysique, UMR Astroparticules et Cosmologie\\
        CEA Saclay\\
        France\\
        E-mail: \email{tagger@cea.fr}}

\abstract{In this contribution I review recent work that goes beyond our model for the Low-Frequency Quasi-Periodic Oscillation of microquasars, based on the Accretion-Ejection Instability. I show that similar instabilities, which can be viewed as strongly unstable versions of the diskoseismologic modes, provide explanations for both the High-Frequency QPO and for the quasi-periodicity observed durng the flares of Sgr A*, the supermassive black hole at the Galactic Center.}

\FullConference{VI Microquasar Workshop: Microquasars and Beyond\\
		September 18-22 2006\\
		Societ\`a del Casino, Como, Italy}

\begin{document}

\section{Introduction}
Explaining the Quasi-Periodic Oscillations (QPO) observed in X-ray binaries has proven to be a challenging attempt. Two of their properties stand out as basic requirements for any model: they 
are {\em coherent}, \ie\ they produce narrow features in the temporal spectrum, and they often have {\em strong amplitudes}. This means first that their explanation must involve a globaly ordered motion of the gas (or at least a global order in their emission pattern), and second that the resulting pattern must be of large amplitude: indeed the low-frequency QPO in microquasars can reach up to 40\% in rms. Thus the mechanism which excites the QPO is as important to identify as the one which determines their frequency. \\
In recent years we have presented, as a possible explanation for the low-frequency QPO (LF-QPO) of microquasars, an MHD instability we called the Accretion-Ejection Instability (AEI), for reasons we will detail below \cite{TP99, RVT02, VRT02, TVRP04}. It occurs in the inner region of accretion disks threaded by a vertical magnetic field of the order of equipartition, \ie\ the configuration favored by MHD models of jets. In a series of publications we have built up the case for this explanation, by further advances in the theory, by numerical simulations and by detailed comparison between observations and the predicted properties of the instability. The main elements in favor of our model are that 
\begin{itemize}
\item the instability forms a {\em global mode} (more properly called a normal mode), \ie\  a coherent pattern rotating in the disk at a single frequency; in fact it shares much of its physics with the self-gravity (rather than MHD)-driven spiral instability of galaxies, which have been described as forming a {\em Quasi-Stationary Spiral Structure}. 
\item Its frequencyis a fraction of the rotation frequency at the inner edge of the disk. The correlation of the mode frequency with the disk inner radius $r_{i}$ changes sign when $r_{i}$ gets close to the Marginally Stable Orbit (MSO), as seems to be seen in observationns.
\item It is an {\em instability}, \ie\ it grows spontaneously to high amplitude without need for an external excitation, if the instability criterion (which depends on the radial of a certain quantity discussed below) is fulfilled.
\item It has the property of extracting gravitational energy and angular momentum from the disk (causing accretion) and redirecting a fraction of it to the corona, where it might energize a jet.
\item Both this and the rest of the accretion energy are evacuated from the disk as waves, rather than dissipated locally, as assumed in \eg\ the $\alpha$-disk model. This implies that the turbulent heating of the disk stops while the corona becomes active, in agreement with the fact that the LF-QPO is associated with the low-hard state of the sources.
\end{itemize}
We have then attempted, assuming that the AEI was indeed the source of the LF-QPO,  to derive a more general understanding of the physics of microquasars and of their variability. For this we turned to \GRS\ and in particular to its $\beta$ class of variability, which has been extensively studied in X rays as well as Infrared and radio. Comparing the behaviour of the source with the physics of the instability, we have derived a {\em Magnetic Floods} scenario in which we are led to attribute the cycles of \GRS\ to the cycling of vertical magnetic flux, advected with the gas and destroyed by reconnection (corresponding to the relativistic ejections) when the conditions permit it.\\
In this contribution I will summarize recent developments which permit us to present possible explanations for both the quasi-periodicity observed during the flares of Sgr A*, the supermassive black hole at the center of the Galaxy, and for the high-frequency QPO (HF-QPO) of microquasars, with its puzzling frequency ratio of 2 to 3 and sometimes higher multiples. I will conclude by discussing how these results may show us how to generalize the Magnetic Floods scenario to the variability of microquasars, in particular to the universal track they seem to follow in Hardness-Intensity Diagrams.
\section{AEI and the Rossby-Wave Instability}
As explained in TP99, the AEI is constrained to occur near the inner edge of the disk, because its physics rests on a reflection of waves there. This is similar to the closely related spiral instability of galactic disks and to the Papaloizou-Pringle instability \cite{PP85, PP87, NGG87}, which represent the same normal modes driven respectively by self-gravity and by pressure. It also depends on a positive gradient of a quantity
\begin{equation}
{\cal L}_{B}=\frac{\kappa^{2}}{2\Omega} \frac{\Sigma}{B^{2}}
\label{eq:lb}
\end{equation}
where $\Omega$ is the rotation frequency in the disk, $\kappa$ the epicyclic frequency (given by $\kappa^{2}=4\Omega^{2}+2\Omega\Omega'r$), $\Sigma$ the surface density, and $B$ the magnetic field threading the disk. The AEI is formed of spiral density waves and Rossby waves (waves that propagate in vorticity gradients), coupled by differential rotation, and 
\lb\ is the parameter that controls the propagation of Rossby waves: the frequency of a wave defines a corotation radius $r_{corot}$, where $\omega=m\Omega$, and Rossby waves can propagate only on one side or the other of $r_{corot}$, depending on the sign of the gradient of \lb. As a result, the flux of energy and angular momentum of a Rossby wave has the same sign as this gradient.\\
On the other hand spiral waves can propagate on both sides of $r_{corot}$; they have a positive energy flux where they rotate faster than the gas ($\omega>m\Omega$, where $m$ is the azimuthal wavenumber), \ie\  beyond  $r_{corot}$, and negative energy in the opposite case, and this is the key to the amplification process: if two waves of opposite energy can couple, they can both grow by exchanging energy and angular momentum. \\
In the AEI $r_{corot}$  is typically at $\sim$ 2 to 3 times the inner disk radius. Spiral waves inside corotation have negative energy, \ie\  they extract energy from the disk, causing accretion, and transfer it to a positive energy Rossby wave which propagates just beyond $r_{corot}$ if the radial gradient of \lb\ is positive. A fraction of this energy can then be re-emitted upward to the corona as an \alf wave. The whole pattern forms a {\em normal mode}, i.e. a standing pattern rotating at a single frequency and growing exponentially to high amplitude, providing the base of our model for the LF-QPO. Let us mention that the associated \alf wave may well have been recently observed, in a different context: a double helix nebula, seen in Spitzer observations of the Galactic Center region, has been interpreted as an \alf wave, presumably emitted from the 2-armed spiral (likely to be self-gravity driven) in  the circumnuclear disk at $\sim$ 1 pc from the Galactic Center \cite{MUD06}. 
\\
When \lb\ has an extremum at a given radius in the disk, another possibility exists to form normal modes (this condition is similar to the one for tthe Kelvin-Helmholtz instablity in classical hydrodynamics): in this case Rossby waves can propagate on both sides of corotation (the details are different if \lb\ has a minimum or a maximum, but we won't discuss this here).  A standing pattern of Rossby waves of opposite energies can thus be formed, and become unstable: this exists in unmagnetized disks \cite{LLC99,LFL00,  LCW01}, where it has been called the Rossby-Wave Instability (RWI), and we can expect it to become more unstable with the help of the Lorentz force in MHD. The question is now of course to find mechanisms that can cause such an extremum. This can result from an extremum of $\Sigma/B$ in equation \ref{eq:lb}, if matter or magnetic flux pile up somewhere in the disk; but it can also come from the term $\kappa^{2}/2\Omega$, \ie\ from the rotation curve.\\
We have recently found three astrophysically relevant cases where this instability may turn out to play an important role:
\begin{itemize}
\item Since its discovery, the Magneto-Rotational Instability (MRI) is widely considered as the main cause of turbulent accretion in disks. However  a difficulty arises in  protostellar accretion disks, where models lead to conclude that the ionization ratio is too low in a `dead zone' (DZ), typically between 1 and 5 AU from the star, so that the MRI cannot act. We have shown \cite{VT06} that lower accretion in the DZ self-consistently causes the gas to pile up at its edge, creaing an extremum of \lb: the RWI can thus act and extend across the DZ to cause accretion. Furthermore, the vortices created by the standing Rossby waves are favoured sites for planet formation.
\item The black hole at the center of the Galaxy is believed to accrete gas mostly in the form of low angular momentum blobs produced by colliding stellar winds. These blobs are first captured in the very faint accretion disk of Sgr A* where they circularize at a few tens of $r_{G}$, and produce flares observed in IR and X-rays. These flares are short (2 to 3 hours) and show a distinct quasi-periodicity at a frequency of the order of the MSO rotation frequency around a 3.6 million $M_{\odot}$ black hole. We have shown \cite{TM06} that, as the circularized blob creates an extremum of density and thus of \lb, the development of an MHD form of the RWI could cause accretion on the required time scale with a strong quasi-periodicity in the required frequency range.
\item As shown is diskoseismologic models, the relativistic rotation curve creates near the MSO the possibility to trap modes (called g-modes in this context): indeed by definition $\kappa$ vanishes at the MSO (inward from the MSO $\kappa$ is imaginary, making circular orbits unstable); on the other hand at large radii, where the rotation is keplerian, $\kappa=\Omega$ and thus decreases with $r$. Thus $\kappa^{2}/2\Omega$ has an extremum slightly beyond the MSO, and so does \lb\ unless the radial profile of $\Sigma/B^{2}$ is unphysically tailored. However it remained unnoticed in  diskoseismologic models that these modes are strongly unstable by the RWI mechanism,  especially if the disk is magnetized. We have found \cite{VT06} that this (or rather a hybrid of the AEI and RWI) provides a possible explanation for the HF-QPO.
\end{itemize}
I detail these results in the next sections.
\section{The flares in Sgr A*}
\label{sec:sgr}
Detailed modelling of the environment of the black hole in Sgr A* has shown that blobs of low angular momentum gas `rain onto' the disk and must circularize at a few tens of gravitational radii. This gas must be magnetized, since it is observed by its synchro-Compton emission. 

\begin{figure}[t] 
\centering
\includegraphics[width=4.25in, trim=50 20 110 0]{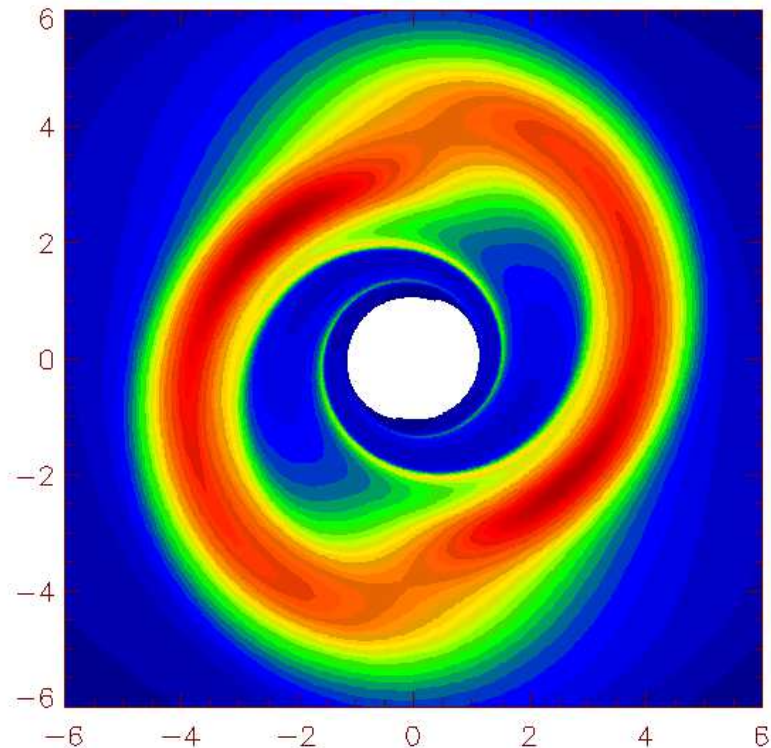} 
   \caption{Surface density in our simulation of Galactic Center flares, when the bump at 
$4\ r_{MSO}$ (red, dark) starts being disrupted by the instability.}
 \label{fig:rho40}
\end{figure}
\begin{figure}[top] 
\centering
\includegraphics[width=.49\textwidth]{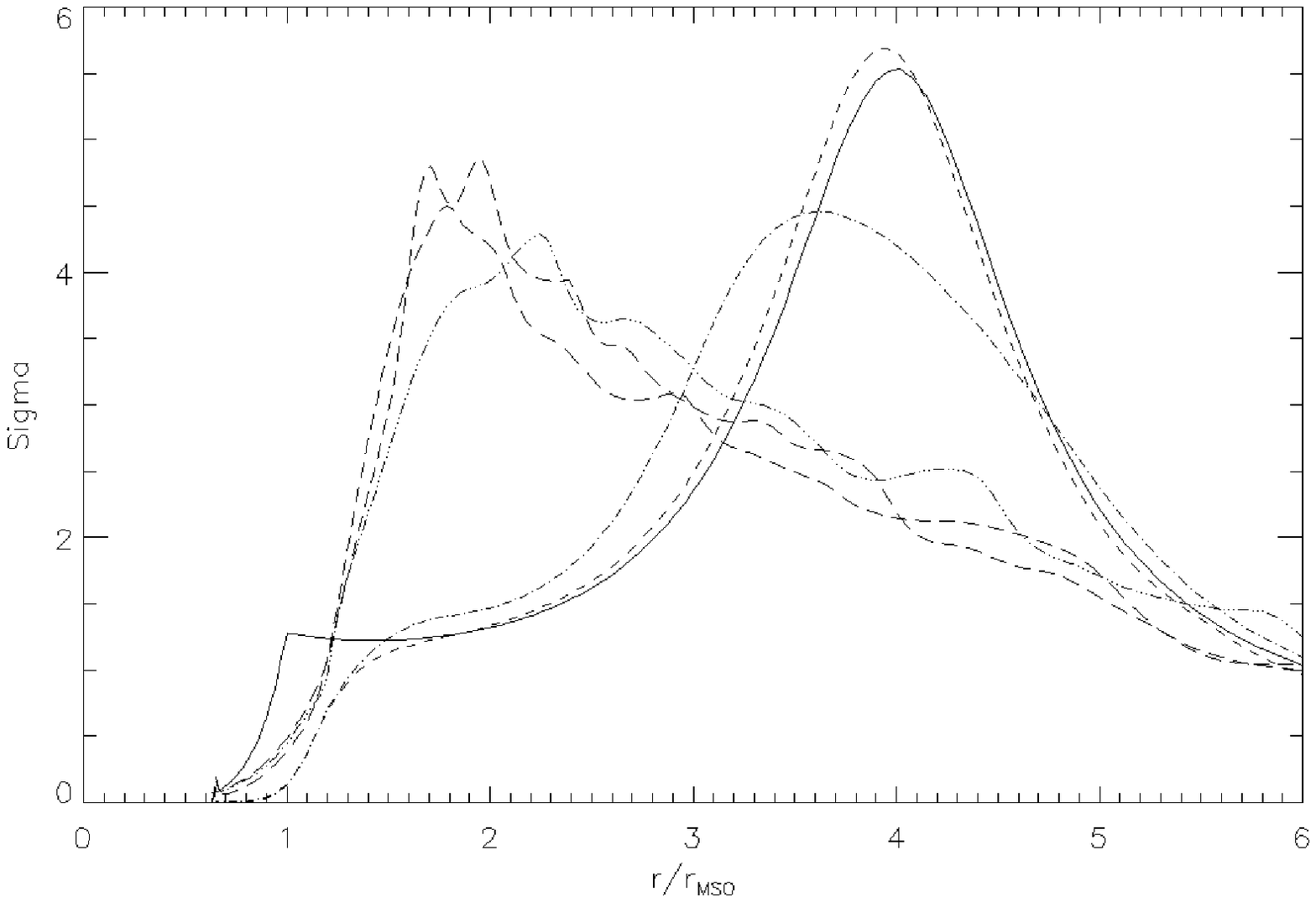} 
\includegraphics[width=.49\textwidth]{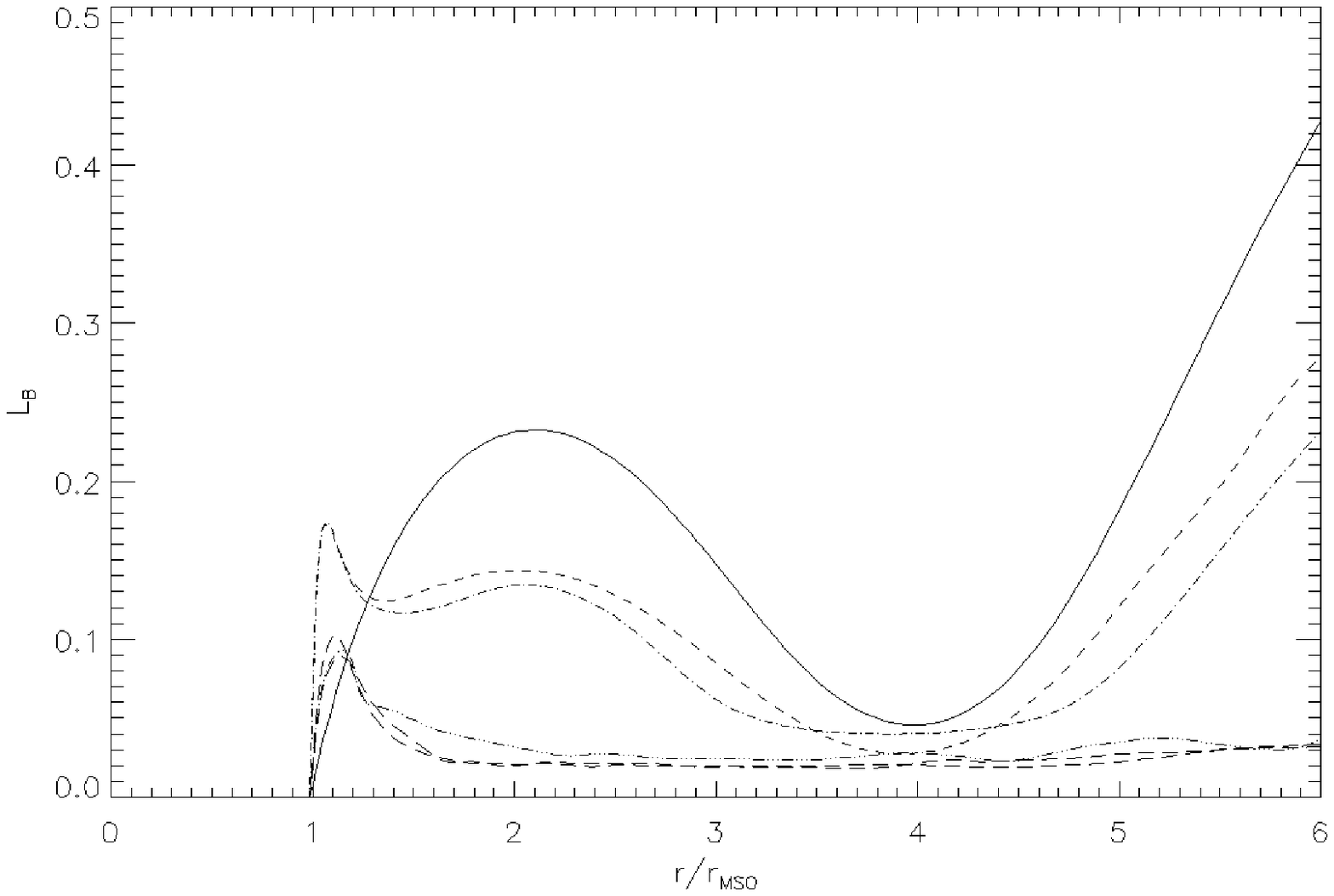} 
\caption{Radial profiles of the surface density (top) and \lb (bottom), at startup (full) and after 20 , 40, 47, 60, 75  MSO orbital times during our simulation.}
   \label{fig:bump4Sigma}
\end{figure}
We have modelled its evolution using the nonlinear MHD numerical code previously used to study the LF-QPO \cite{CT01}. This code describes an infinitely thin disk in vacuum, a constraint that prevents us from studying the MRI (which varies across the disk thickness) but is well adapted to the physics of the spiral and Rossby waves considered here. We have modified this code by introducing a Paczinsky-Witta pseudo-Newtonian potential 
\(
\Phi(r)={GM}/({r-r_{S}}).
\)
which mimicks the relativistic rotation curve and creates a MSO at $3r_{S}$, where $r_{S}$ is the Schwartzschild radius. The initial conditions are axisymmetric, plus a very low amplitude noise from which the instability will grow. The radial profile represents a faint disk, to which is superimposed a circularized blob centered at 4 $r_{MSO}$. The simulation extends from $r\approx .6$ to $15 r_{MSO}$. At the inner boundary the gas is allowed to flow freely (stress-free boundary condition) toward the black hole.\\
In a very short time the disk forms self-consistently a plunging region near the MSO, where the gas is accelerated to supersonic speed and disappears at the inner boundary. A strong 2-armed spiral wave is then observed to grow rapidly inward from the blob: this is the spiral wave emitted by the Rossby waves that develop in the circularized blob, and rapidly cause it to accrete to the black hole. Figure~\ref{fig:rho40} shows the surface density profile at $t=40t_{MSO}$, 
when the density bump starts being disrupted by the instability. Figure~\ref{fig:bump4Sigma} shows 
that after this rapid phase, the radial profile of \lb becomes perfectly flat beyond the MSO region,  
confirming qualitatively that it is that profile to which the instability reacts.
\begin{figure}[t] 
\centering
\includegraphics[width=.6\textwidth]{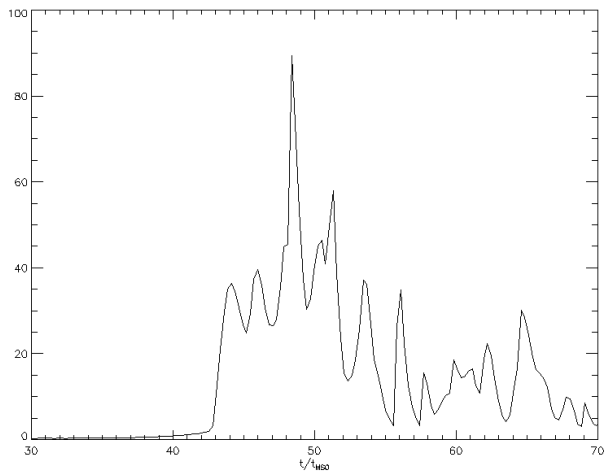} 
   \caption{Accretion rate through the inner edge of the simulation box, as a function of time (in units of the rotation time at the MSO).}
   \label{fig:lightcurve}
\end{figure}
 Figure \ref{fig:lightcurve} shows the resulting accretion rate measured at the MSO. Both the duration of the flare ($\sim$15 rotation times) and the quasi-period ($\sim$1.5 rotation times) are compatible with the observed ones. One should note that the quasi-period should be sensitive to the details of the rotation curve, fully relativistic or pseudo-Newtonian, so that using it to estimate the spin of the black hole is not a straightforward process.
\section{High-Frequency QPO}
\begin{figure}[t] 
\centering
\includegraphics[width=.6\textwidth]{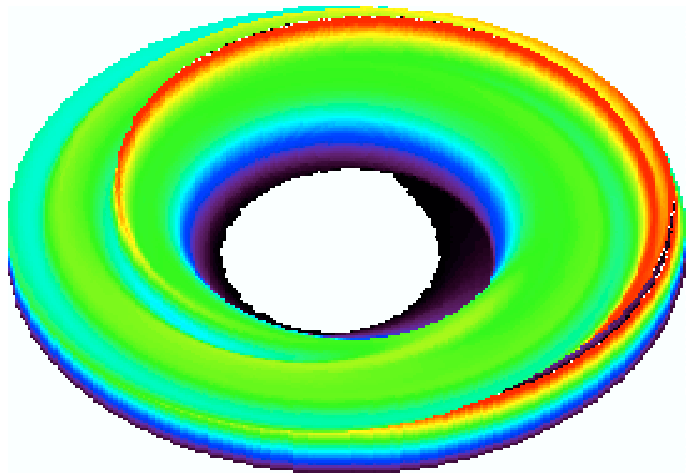} 
   \caption{Surface density profile in our first simulation of HF-QPO, with a stress-free  boundary condition.}
   \label{fig:sigmaHF1}
\end{figure}

\begin{figure}[t] 
\includegraphics[width=.49\textwidth]{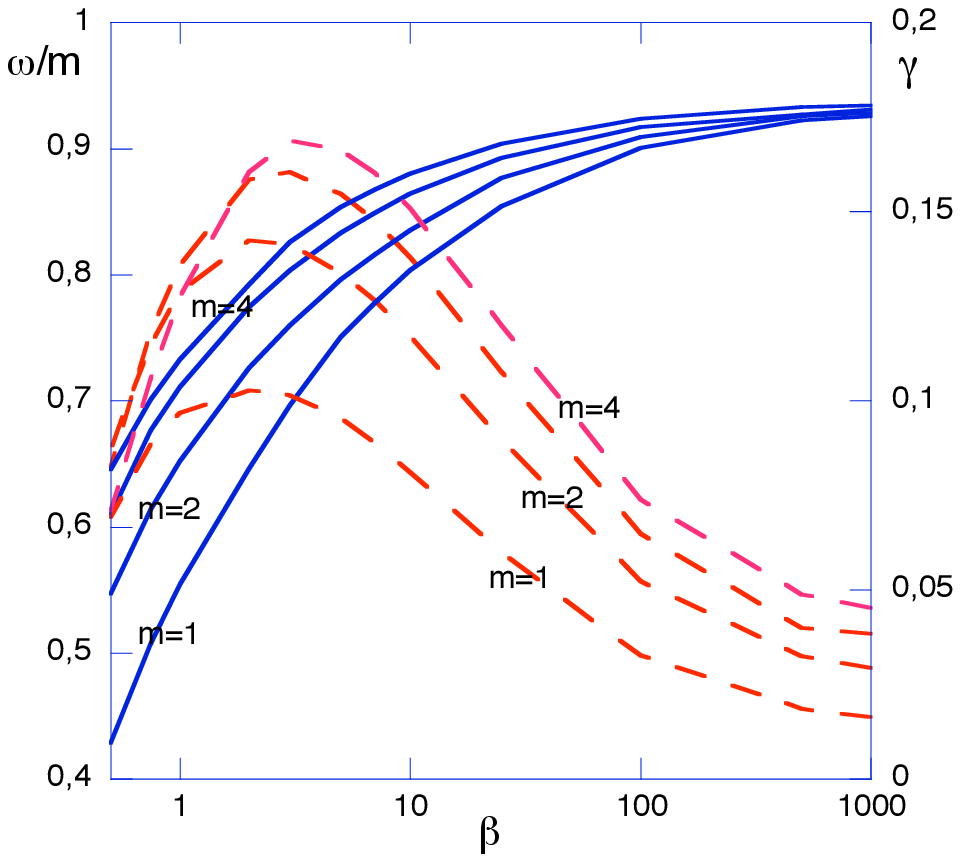} 
\includegraphics[width=.49\textwidth]{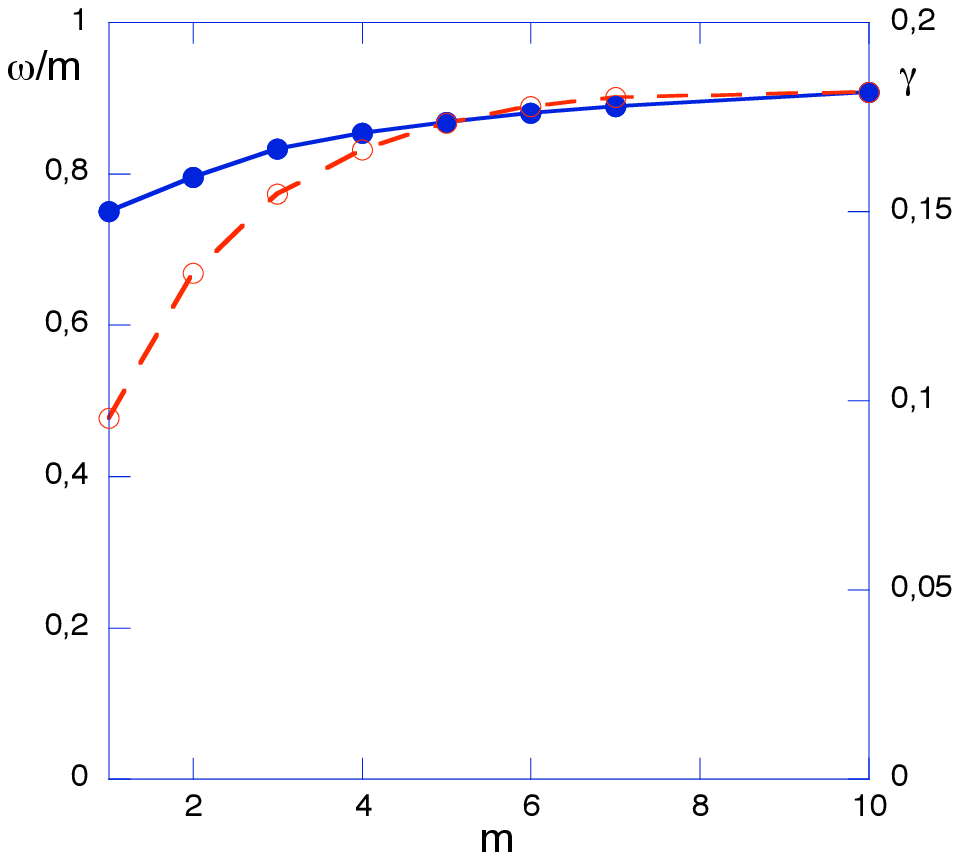} 
\caption{Pattern speed $\omega/m$  (solid) and growth rate $\gamma$ (dashed) of the linear eigenmodes. Top, as a function of $\beta=8\pi p/B^{2}$; bottom, as a function of $m$ for $\beta=5$. The results are normalized to the rotation frequency at the MSO.}
\label{fig:linear}
\end{figure}

\begin{figure}[t] 
\centering
\includegraphics[width=.6\textwidth]{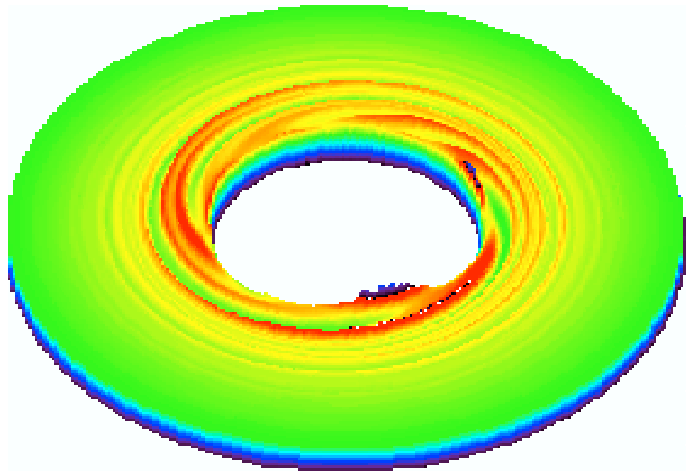} 
   \caption{Surface density profile in our second simulation of HF-QPO, with a reflecting boundary condition, showing a dominant $m=3$ mode.}
   \label{fig:sigmaHF2}
\end{figure}

Figure \ref{fig:bump4Sigma} illustrates the fact that, as mentioned in our introduction, the relativistic rotation curve self-consistently creates an extremum of \lb\ just beyond the MSO and thus a different way to make the RWI unstable. When the inner edge of the disk reaches down to the MSO, it should thus be unstable by this mechanism, and it is tempting to seek there an explanation for the HF-QPO.  \\
We have first run simulations using the same setup as in section \ref{sec:sgr} without the added blob density, \ie\ with a smooth density and magnetic field profile, and stress-free boundary condition. Figure \ref{fig:sigmaHF1} shows that we find a strong one-armed instability, which slowly erodes the self-consistently established plunging region. Varying the initial parameters and profiles always shows that the disk selects an $m=1$ mode. Thus, although the result seems generic to this configuration, and thus well adapted to explain the HF-QPO of microquasars, it gives no indication that could explain the surprising observation that these QPO are seen at commensurable frequencies at 2,3 and sometimes more times a fundamental which is never observed.
We have then turned to a different inner boundary condition, and find that it yields a much more interesting result: we use at the inner boundary a reflecting condition, by fixing $u_{r}=0$ there. Let us first describe this result, before we discuss the relevance of this boundary condition.\\
We have analyzed this both by direct solution of the linear eigenmode equations and by numerical simulations. Figure \ref{fig:linear} shows that the modes have very nearly the same pattern speed, $\omega/m$, very close to the rotation frequency at the MSO, because they are constrained to have their corotation radius in the local extremum of \lb\ due to the relativistic rotation curve. These modes will thus appear very near multiples of a fundamental frequency.
Figure \ref{fig:linear} also shows that the growth rates of these modes ares very similar, except for the $m=1$ which has a significantly lower one: one can thus expect that, depending on the conditions and on non-linear effects, any of these modes should dominate, except for the $m=1$. This expectation is borne out by the nonlinear simulations where, as illustrated in figure \ref{fig:sigmaHF2}, we find a dominant $m=3$ mode. Varying the conditions in the simulations we can obtain different values for the dominant $m$, but never $m=1$ unless we arbitrarily suppress any other seed nonaxisymmetric perturbation initially. This thus seems to present a good explanation for the observations of the HF-QPO. 
\\
Let us now discuss the boundary condition: let us first note that the stress-free boundary condition, which is commonly used, is a legacy from previous hydrodynamical studies. But using it in MHD lets the {\em magnetic flux}, as well as the gas, flow freely toward the black hole! This, of course, cannot be right. Unless the gas sheds all of its magnetic flux in the disk before it crosses the MSO, this flux must accumulate in the cavity between the disk and the black hole. The horizontal part of this flux can be expected to be evacuated vertically by buoyancy, but this does not apply to the vertical flux. Thus using the stress-free condition may be correct for simulations of very short events, such as the flares of Sgr A* studied in section \ref{sec:sgr}. In that case one can expect that the flux accumulated during a flare may re-expand radially in the period between the flares, where the accretion rate is extremely weak. But this can certainly not apply to a model of HF-QPO, which are usually observed during long stretches of the Steep-Power Law (SPL) state of microquasars. The vertical flux must thus accumulate between the black hole and the disk, and our hypothesis is that the magnetic structure holding this flux, while it still permits accretion on the slow diffusive time scale of accretion, acts as a reflecting boundary on the fast time scale of the waves we consider. As a final argument let us add that, just as the AEI, the Rossby wave component of the instability will radiate energy to the corona as \alf waves: this would explain that the SPL state is characterized by a strong high-energy emission, presumably of inverse Compton origin, and a strong disk emission. We believe that this can correspond to our model, where the instability is very localized near the MSO so that, contrary to the low state, the turbulently heated disk can still be observed while the comptonized emission is seen.\\
\section{Discussion}
Our Magnetic Floods scenario (MFS) had led us to consider the interplay between the vertical magnetic flux advected with the gas in the disk and the flux stored in the cavity between the disk and the black hole as the key to understand the variability of microquasars, in particular \GRS, for two reasons:
\begin{itemize}
\item the first one is that,  if one believes that the AEI is indeed the cause of the LF-QPO, one has to seek an explanation for the transitions between the `thermal dominant' (or high-soft) state and the low state, which exhibits the LF-QPO: the most natural explanation is to identify these states as dominated respectively by MRI turbulence and by the AEI; this in turn can be explained by the accumulation of magnetic flux in the inner disk region, causing a transition from $\beta>1$, where the MRI dominates to $\beta<1$, where the MRI is stabilized and the AEI is strong. This view is supported in particular by observations where the LF-QPO seems to appear {\em before} the state transition, so that it would be the cause, rather than the consequence, of this transition. 
\item the second reason is that if, as commonly believed, the ejection marking the end of the low state is due to a reconnection event, this requires that regions carrying opposite magnetic fluxes be brought in contact. In the magnetic geometry we use, which is also that of MHD models of jets, where a vertical magnetic field threads the disk, this would imply that the magnetic flux in the disk and the flux stored in the central cavity between the disk and the black hole (some of the latter threading the black hole, \cite{BZ77}) have opposite polarities. We have argued that this could explain the `forbidden' transition betw\`een states C and B, as identified and discussed in \cite{Bel00}: once in state C (the low-hard state with the LF-QPO, more magnetized in our interpretation), the source can change states only by destroying magnetic flux, \ie\ by reconnection, thus returning to the `thermal dominant', less magnetized state A.
\end{itemize}
Thus in our interpretation the classes of variability of \GRS\ that show cycles including state C and an ejection can occur only in situations where the magnetic fluxes in the disk and in the central cavity are antiparallel. Let us here consider the analogy with the interface between the solar wind and the Earth magnetosphere: at this interface one can also have either parallel or antiparallel magnetic fields, and only the latter configuration is prone to reconnection. This analogy leads us to suggest that the dichotomy observed in \GRS\ between variability classes that exhibit state C and ejections, and classes that exhibit only states A and B, may correspond to a dichotomy between situations where the magnetic fluxes in the disk and the central cavity are parallel or antiparallel.\\
Our new result on the HF-QPO adds a new argument to this discussion: we have shown that, if something (for which a magnetic structure of the nature we have discussed would be the best candidate) acts as a reflecting boundary for waves, then there is a strongly unstable global mode, inherent to the relativistic rotation curve when the disk extends down to its MSO, which has all the properties to explain the HF-QPO and their harmonic ratio. Thus once again, and for totally different reasons, we are led to consider the magnetic flux stored in the central cavity as the key parameter that controls the variability of the source.\\
Could this scenario be extended to other microquasars? A most useful guide to answer this question may be the universal track, in Hardness-Intensity diagrams, that seems to be followed by all microquasars \cite{HB05}. Guided by the previous discussion, we believe that the hysteresis cycle found in this diagram may be explained by the accumulation of vertical magnetic flux in the central cavity surrounding the black hole, and the advection of additional flux in the disk. Since the latter is probably due to a turbulent dynamo, in the disk itself or the companion star, chaotic field reversals in this dynamo could change the magnetic configuration at the disk/cavity interface from parallel to antiparallel, and after some time cancel the trapped flux. We will discuss this extended scenario in details in a forthcoming publication.

\acknowledgments
It is a pleasure to acknowledge many helpful discussions and much common work, from which the present contribution is derived, with M. Cadolle-Bel, R. Remillard, J. Rodriguez and P. Varni\`ere.

\bibliographystyle{PoS}

\end{document}